\documentclass[a4paper,11pt]{article}
\usepackage{pos}
\usepackage{siunitx}
\usepackage{todonotes}
\usepackage{wrapfig}
\usepackage{hyperref}
\usepackage{subfig}
\renewenvironment{thebibliography}[1]{
  \begin{oldthebibliography}{#1}
    \setlength{\itemsep}{-0.03em}
    \setlength{\parskip}{-0.03em}
}
{
  \end{oldthebibliography}
}

\newcommand{\mycomment}[1]{}

\usepackage{titlesec}
\titlespacing\section{0pt}{4pt plus 4pt minus 2pt}{2pt plus 2pt minus 2pt}

\title{Design, performance and future prospects of vertex detectors at the FCC-ee}

\author*[a]{Armin Ilg}
\author[b]{Fabrizio Palla}

\onbehalf{on behalf of FCC}

\affiliation[a]{Physik Institut, University of Zürich\\
  Winterthurerstrasse 190, 8057 Zürich, Switzerland}

\affiliation[b]{INFN Pisa,\\
Largo Bruno Pontecorvo 3, 56127 Pisa, Italy}

\emailAdd{armin.ilg@cern.ch}
\emailAdd{fabrizio.palla@cern.ch}

\abstract{The CERN proposed $e^+e^-$ Future Circular Collider (FCC-ee) is an electroweak, flavour, Higgs and top factory with unprecedented luminosities. Many measurements at the FCC-ee will rely on precisely determining the particle production vertices using dedicated vertex detectors. 

All vertex detector designs use Monolithic Active Pixel Sensors (MAPS) with a single-hit resolution of $\approx\SI{3}{\mu m}$ and a material budget as low as $\SI{0.3}{\percent}$ of a radiation length per detection layer, which is within specifications for most of the physics analyses.  

This contribution presents the status of the fully engineered vertex detectors and their integration with the collider beam pipe and discusses their predicted performance using the DD4hep full simulation. A concept for an ultra-light vertex detector using curved wafer-scale MAPS is also presented, which allows reducing the material budget by almost a factor of three. This improves the vertexing capabilities, especially for heavy flavour decays, such as $B^0 \rightarrow K^{*0} \tau^+ \tau^-$.
}

\FullConference{%
  42nd International Conference on High Energy physics - ICHEP2024\\
  17-24 August, 2024\\
  Prague, Czech Republic
}

\begin{document}
\maketitle

\section{The Future Circular Collider integrated programme}

The Large Hadron Collider (LHC) at CERN probes fundamental constituents of matter, the elementary particles, and their interactions to ever higher scales in energy and precision. The discovery of the Higgs boson in 2012 completed the Standard Model (SM) of particle physics. Since many open questions of the SM are intertwined with the properties of the Higgs boson, the study of this particle is of great importance at the current LHC and future High-Luminosity LHC (HL-LHC).
However, due to the large number of simultaneous collisions and the QCD nature of hadron collisions, the precise study of the Higgs boson is challenging at the (HL-)LHC.


The Future Circular Collider (FCC) project foresees a new \SI{90.7}{km} circumference collider at CERN to serve the HEP community for the rest of the 21st century. The FCC-ee \cite{2019_CDR_lepton_collider} is planned to start operation around 2045 and would collide intense $e^-$ and $e^+$ beams with unprecedented luminosities. The foreseen centre-of-mass energies ($\sqrt{s}$) from $\sim 90$ up to \SI{365}{GeV} and high instantaneous luminosities of up to $\SI{140e34}{cm^{-2}s^{-1}}$ make it not only a Higgs but also an electroweak, flavour and top factory. The FCC-ee will feature four interaction points, and the tunnel and further infrastructure are made compatible with a later FCC-hh \cite{fcc_hh_CDR} providing hadron collisions at around $\sqrt{s} = \SI{100}{TeV}$. 
The FCC feasibility study started in 2021 and will conclude by March 2025, giving input to the next update of the European Particle Physics Strategy in 2025--2026.


The systematic uncertainties must match the tiny statistical uncertainties to fully benefit from the tremendous number of clean collisions at FCC-ee. Three detector concepts tackling this are under study: IDEA \cite{IDEA}, CLD \cite{CLD}, and ALLEGRO \cite{ALLEGRO}. At their centres, vertex detectors (VXDs) reconstruct the locations of the primary interaction, and secondary and tertiary decay vertices. 

\section{Tracking and vertexing at FCC-ee}

Efficient and precise vertex reconstruction is crucial for flavour tagging and particle lifetime measurements. Therefore, it is essential for many aspects of the FCC-ee physics programme, such as measuring the Higgs coupling to the second-generation quarks and the tau and reconstructing complex flavour physics decay chains.

The vertexing performance is determined by the spatial resolution of the sensors, their distance to the interaction point, the number of recorded hits and the number of radiation lengths ($X_0$) of the beam pipe and the VXD, commonly referred to as \textit{material budget}. A useful metric to characterise the vertexing performance of a given detector design is the resolution of the transverse impact parameter $\sigma_{d_0}$, which can be parametrised as $a \oplus b/(p \sin^{3/2} \theta)$. The parameter $a$ describes the spatial resolution of the sensors used, while $b$ parametrises the effect of the multiple Coulomb scattering due to the material budget. $\sigma_{d_0}$ also depends on the momentum $p$ of the particle and the angle $\theta$ with respect to the beam direction. For FCC-ee, the target is $a\approx \SI{3}{\mu m}$ and $b\approx \SI{15}{\mu m GeV}$.

CLD foresees three barrel layers and three disks per side, which are double-sided to provide a total of six hits for each track in the acceptance. The IDEA VXD design (also used for ALLEGRO) will instead use thinner layers that only provide one hit per layer with at least five hits per track. A silicon tracker surrounds the CLD VXD, while IDEA foresees a lightweight drift chamber as tracker. Both VXD designs expect to use Monolithic Active Pixel Sensors (MAPS) thanks to their low material budget, low power consumption and small pixel pitch capability. 



For IDEA, an engineering model of the VXD is being developed \cite{Boscolo2023,fcc-ee_interaction_region_2024} and integrated into the machine-detector interface region \cite{Boscolo2021} in the context of the FCC feasibility study.

\section{IDEA vertex detector design and integration into machine}\label{sec:idea_vertex_design_mdi}

The IDEA VXD comprises three inner vertex barrel layers, the middle and outer vertex barrel layers, and three disks per side. \autoref{tab:vertex} summarises the properties of the IDEA VXD.

\begin{table}[htbp]
    \vspace*{-0.2cm}
    \centering
    \footnotesize
    \begin{tabular}{c|c|c|c|c|c|c}
        Subsystem & Layer ID & $r$ [mm] & $|z|$ [mm] & $\cos(\theta)$ acceptance & \#ladders & \#Modules/ladder \\ \hline \hline
        Inner vertex barrel & 0 & 13.7 & $<96.5$ & $<0.990$ & 15 & 6 \\
        Inner vertex barrel & 1 & 22.7 & $<160.9$ & $<0.990$ & 24 & 10 \\
        Inner vertex barrel & 2 & 34 & $<257.5$ & $<0.991$ & 36 & 16 \\ \hline
        Middle vertex barrel & 3 & 140 & $<163.1$ & $<0.991$ & 23 & 8 \\
        Outer vertex barrel & 4 & 315 & $<326.3$ & $<0.991$ & 51 & 16 \\ \hline

        Disks & 0 & $34.5 < r < 275$ & 279.1 & $0.712 < \theta < 0.992$ & 56 & 2--6 \\
        Disks & 1 & $70 < r < 315$ & 609.1 & $0.902 < \theta < 0.993$ & 48 & 3--7 \\
        Disks & 2 & $105 < r < 315$ & 918.6 & $0.946 < \theta < 0.994$ & 40 & 4--7 \\
    \end{tabular}
    \vspace*{-0.3cm}
    \caption{Layout of the IDEA VXD for FCC-ee.}
    \label{tab:vertex}
    \vspace*{-0.2cm}
\end{table}

\begin{figure}[htbp]
    \vspace*{-0.2cm}
    \centering
    \subfloat[Ladders]{
        \includegraphics[height=0.145\textheight,trim=0cm 0cm 0cm 2cm,clip]{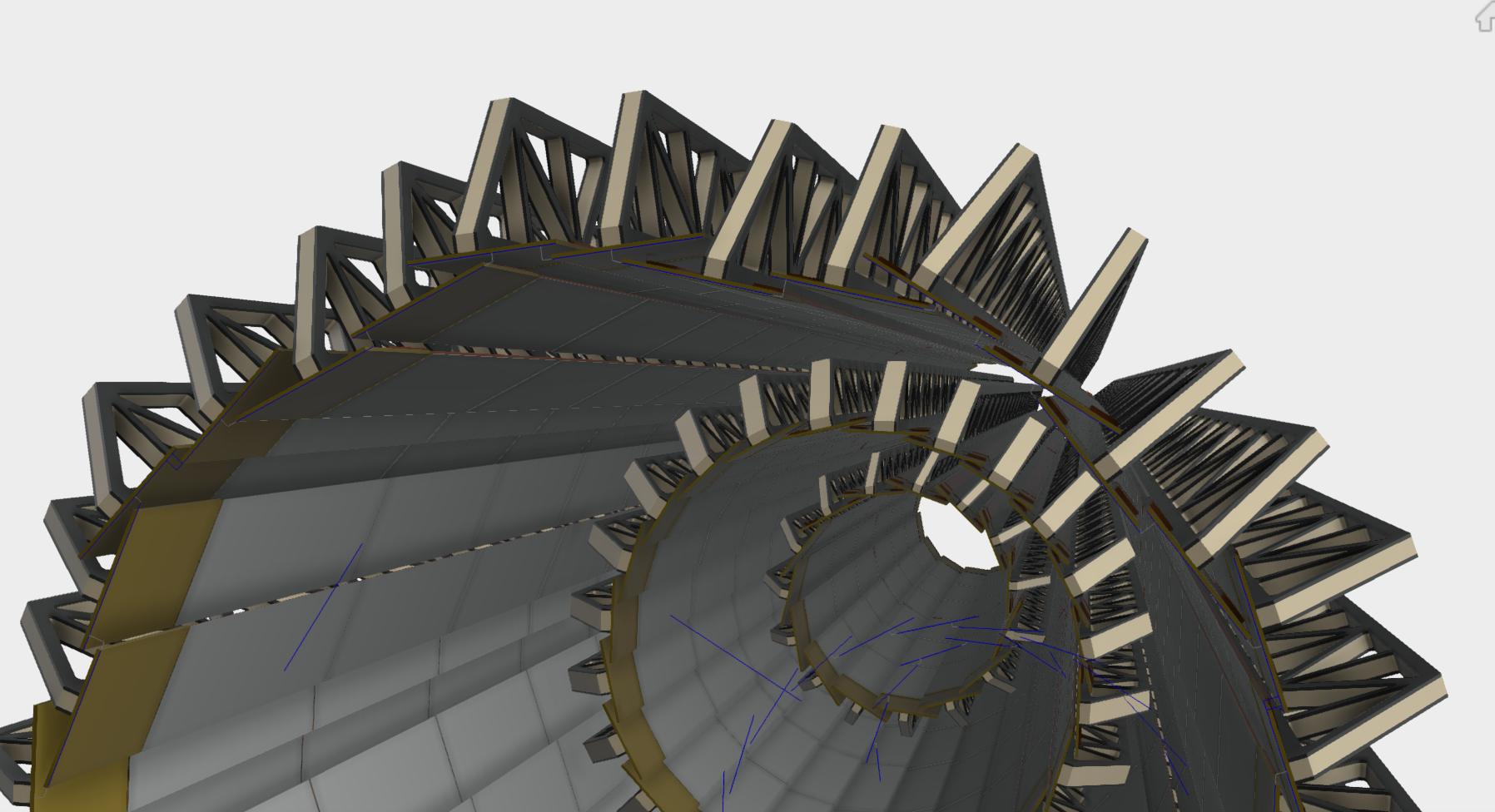}
        \label{fig:idea_engineering_design_inner}
    }
    \hspace*{-0.6cm}
    \subfloat[Longitudinal section]{
        \includegraphics[height=0.145\textheight]{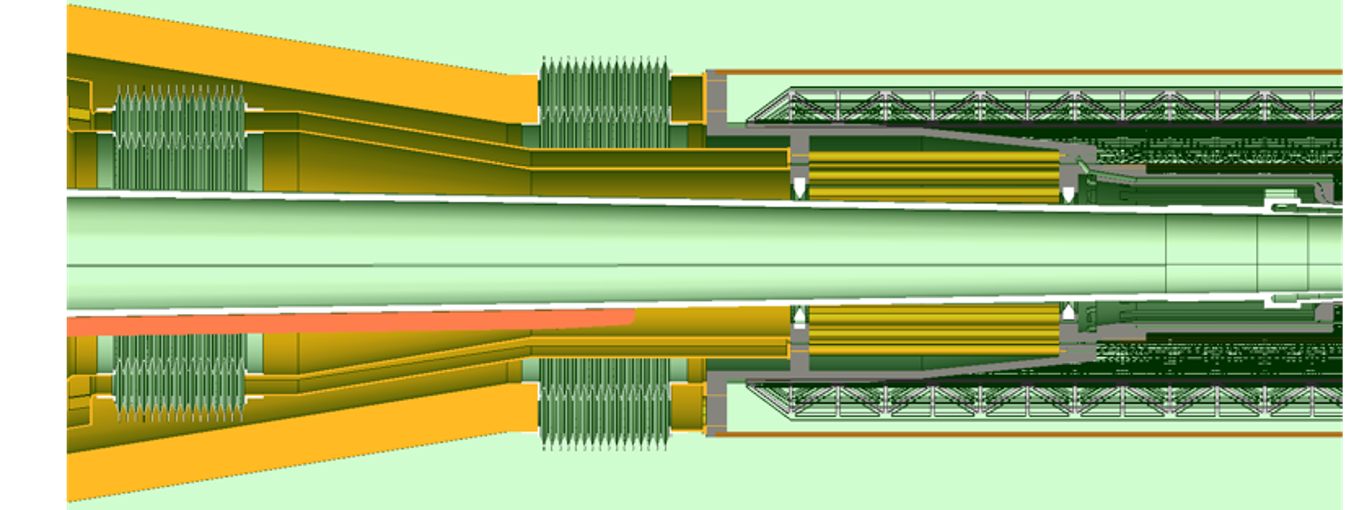}
        \label{fig:cooling_air}
    }
    \vspace*{-0.2cm}
    
    \caption{CAD model of the IDEA inner vertex barrel ladders (a) and longitudinal section including the cooling cones (yellow) and the conical support (grey) on top of the beam pipe (white) (b).}
    \label{fig:}
    \vspace*{-0.1cm}    
\end{figure}

Figure~\autoref{fig:idea_engineering_design_inner} shows the inner vertex barrel CAD design. ARCADIA MAPS \cite{Pancheri2020} with dimensions of $8.4$ ($r$-$\phi$) $\times$ $32$ ($z$) $\SI{}{mm^2}$ are placed along $z$ on top of a lightweight support structure with readout flexes for powering and readout, forming \textit{ladders}. By overlapping these ladders, the insensitive periphery of \SI{2}{mm} in $r$-$\phi$ can be covered to ensure full hermiticity. The ladders are mounted on a conical support made of carbon fibre (CF) that sits on top of the beam pipe and thermally and electrically isolates it from the vertex. The inner vertex is cooled by flowing gas: both air and Helium are being considered. A system of CF cones forces gas convection inside the detector volume and contains power and readout circuits, as shown in Fig.~\ref{fig:cooling_air}. The inner VXD acceptance goes down to $\cos(\theta) < 0.990$, with all support structures and supports clearing the area $\cos(\theta) > 0.994$ to avoid material in front of the luminosity calorimeter (lumical).

\begin{wrapfigure}[9]{r}{0.51\textwidth}
    \centering    
    \vspace*{-0.4cm}
    \includegraphics[width=\linewidth]{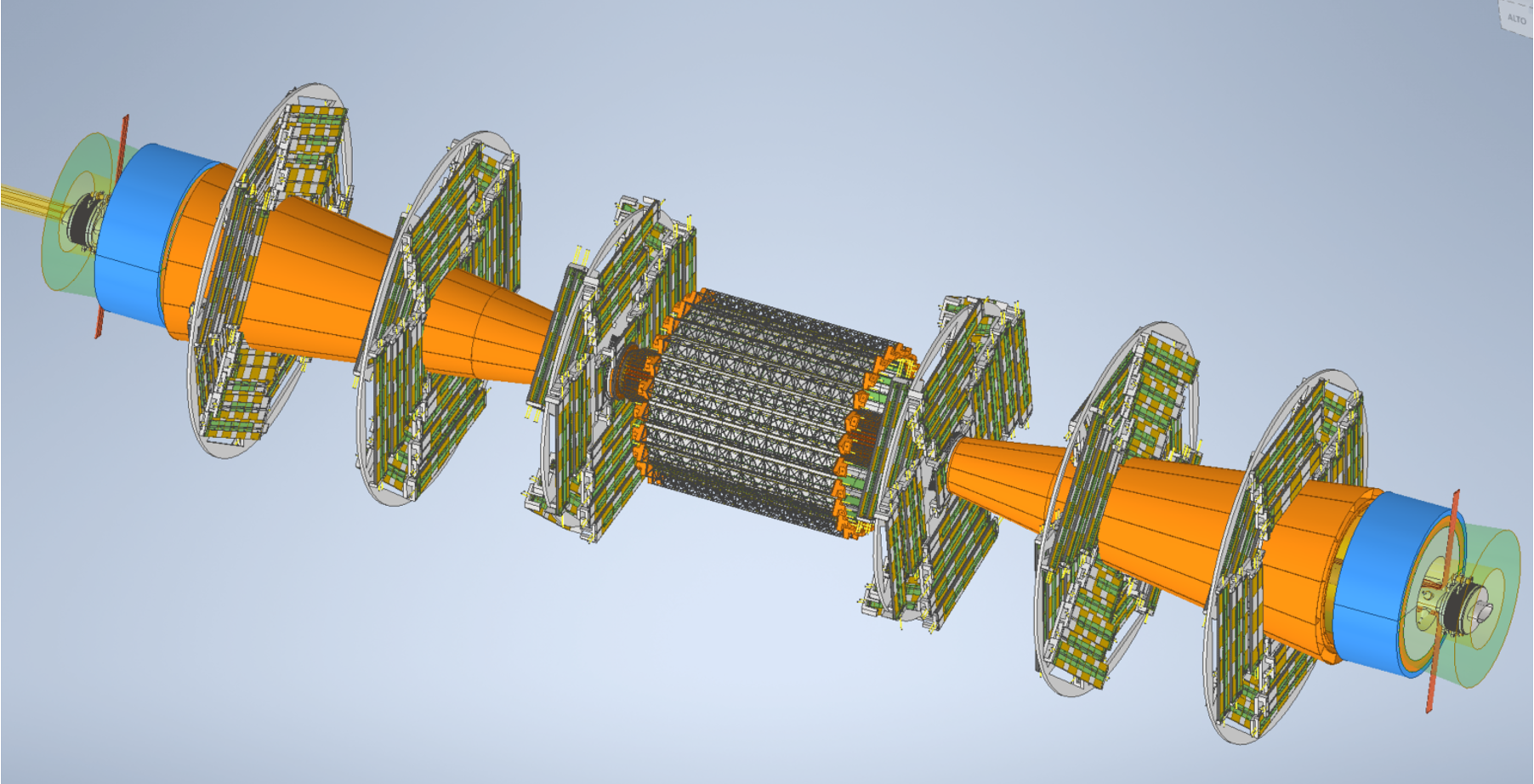}
    \vspace*{-0.6cm}
    \caption{CAD model of the complete IDEA VXD, beam pipe (grey) and the lumical (blue).}
    \label{fig:idea_engineering_design_complete}
\end{wrapfigure}

The cooling performance is assessed by computational fluid dynamics (CFD) simulations, resulting in the largest temperature difference between sensor modules of the third layer of less than $\SI{15}{\celsius}$. All other layers dissipate less power. Mechanical vibration analysis has been performed using a finite element analysis (FEA) in ANSYS, resulting in the magnitude of the maximum displacement of about $\SI{1.5}{\mu m}$ for a nominal airflow of \SI{0.7}{g/s}.

The middle and outer vertex barrel and the disks use quad modules of ATLASPix3 \cite{Peric2021}, which sit on multilayered CF support structures with water cooling pipes in the middle. A lightweight triangular truss structure holds the ladders of the barrels in place. The disk ladders are arranged in petals and attached to supports of CF walls interleaved with Rohacell. Finally, a CF lightweight support structure holds in position the interaction region beam pipe, the VXD, and the lumical \cite{Boscolo2023}. \autoref{fig:idea_engineering_design_complete} displays the complete VXD CAD model. In the following, the IDEA VXD was implemented in DD4hep \cite{Gaede2020} to perform full simulation studies and estimate the vertexing performance.

\section{\mbox{IDEA vertex detector performance in full simulation}}

\begin{wrapfigure}[9]{R}{0.44\textwidth}
    \centering
    \vspace*{-0.4cm}
    \includegraphics[width=\linewidth,trim=1cm 0.5cm 1cm 0.7cm,clip]{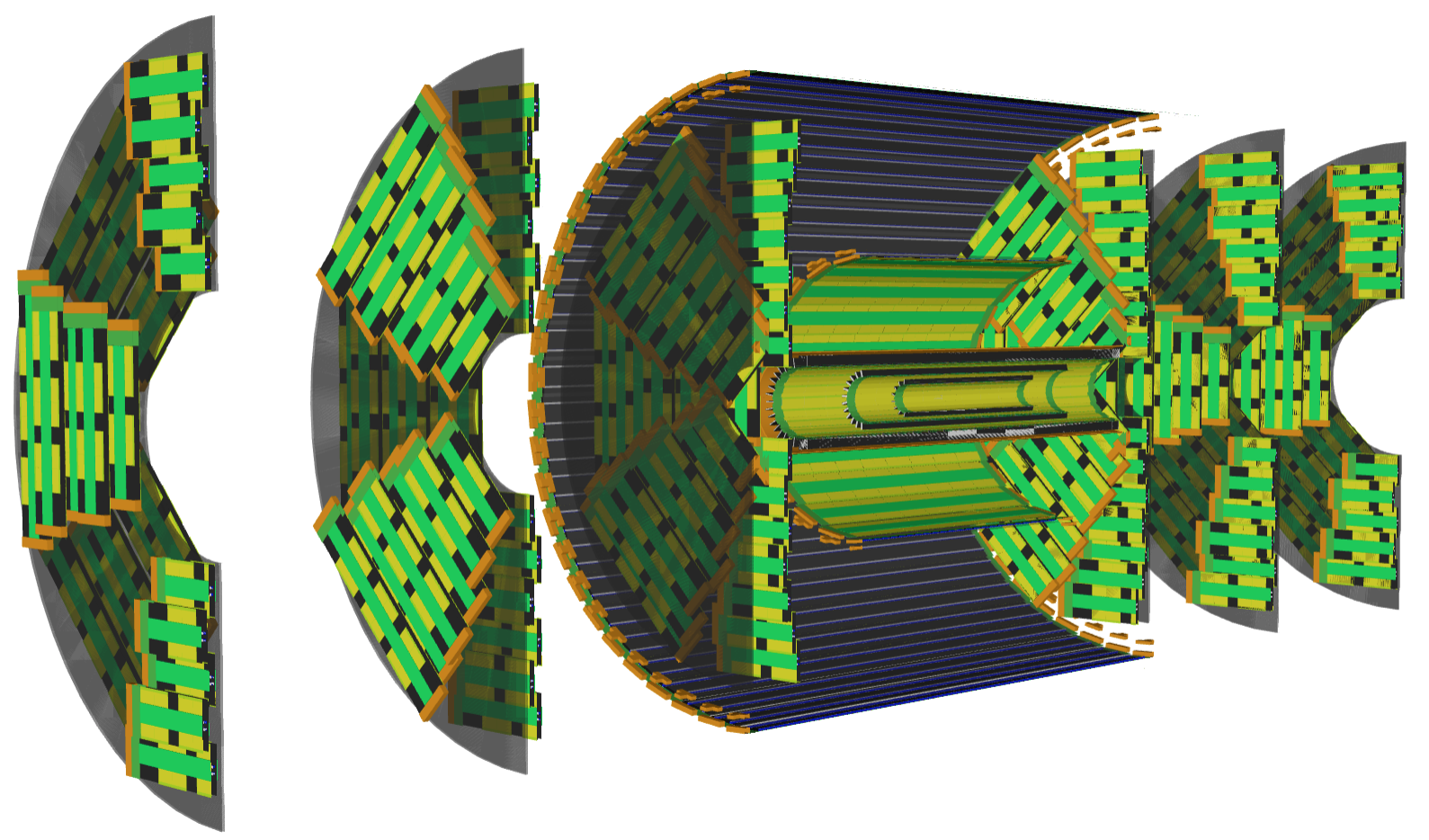}
    \vspace*{-0.6cm}
    \caption{IDEA VXD in DD4hep.}
    \label{fig:vtx_dd4hep}
\end{wrapfigure}

The description of the IDEA VXD in DD4hep (see \autoref{fig:vtx_dd4hep}) closely follows the CAD model. The ladder support and readout flexes, for example, use the correct material stack. The insensitive sensor peripheries are accurately modelled, which will allow for an honest estimation of the VXD hermeticity. Proxy volumes describe the truss and disk support structures with a reduced density to yield the same total mass.

\begin{figure}[htbp]
    \vspace*{-0.4cm}    
    
    \centering
        \subfloat[First inner vertex barrel layer]{
        \includegraphics[width=0.3\textwidth]{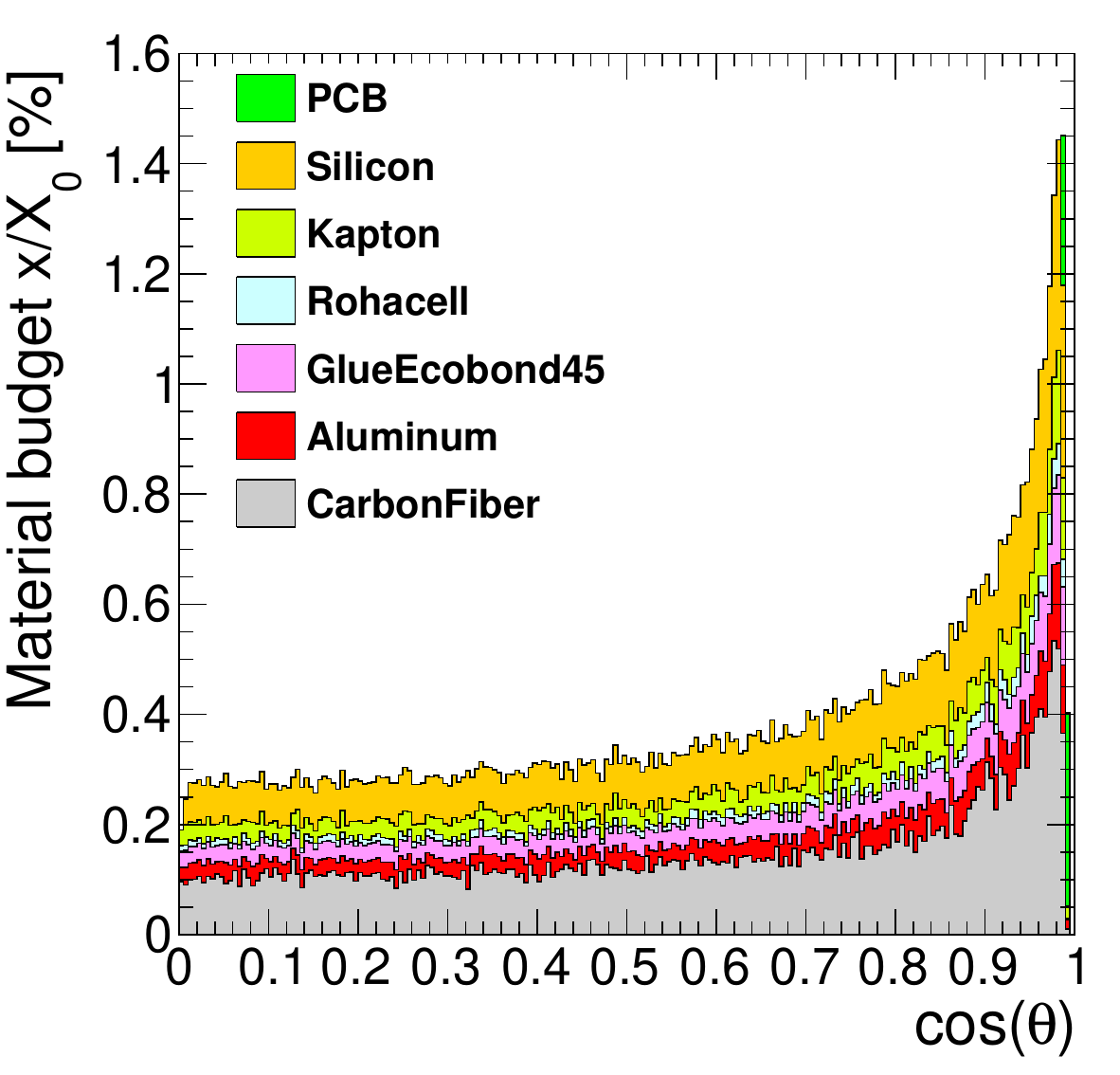}
    }
        \subfloat[First inner vertex barrel layer, $\cos(\theta)$--$\phi$]{
        \includegraphics[width=0.39\textwidth]{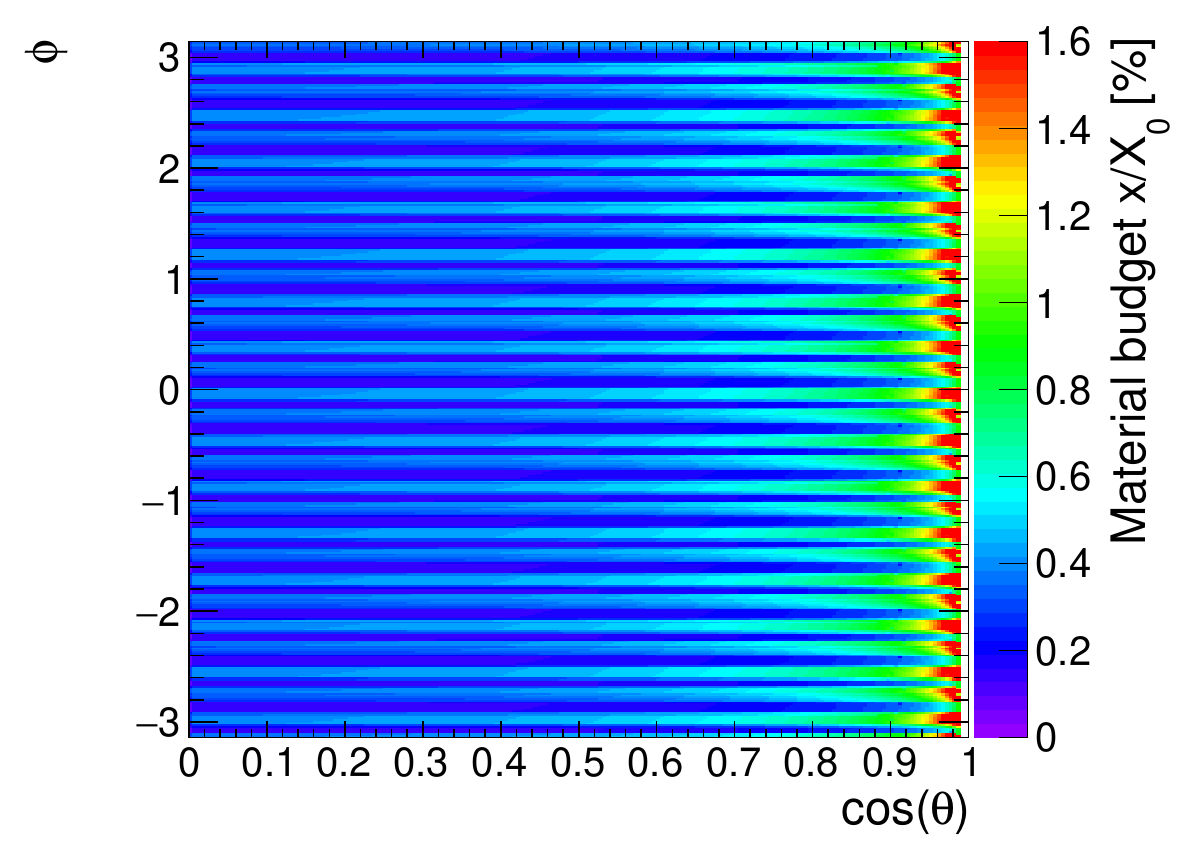}
    }
        \subfloat[Total VXD]{
        \includegraphics[width=0.3\textwidth]{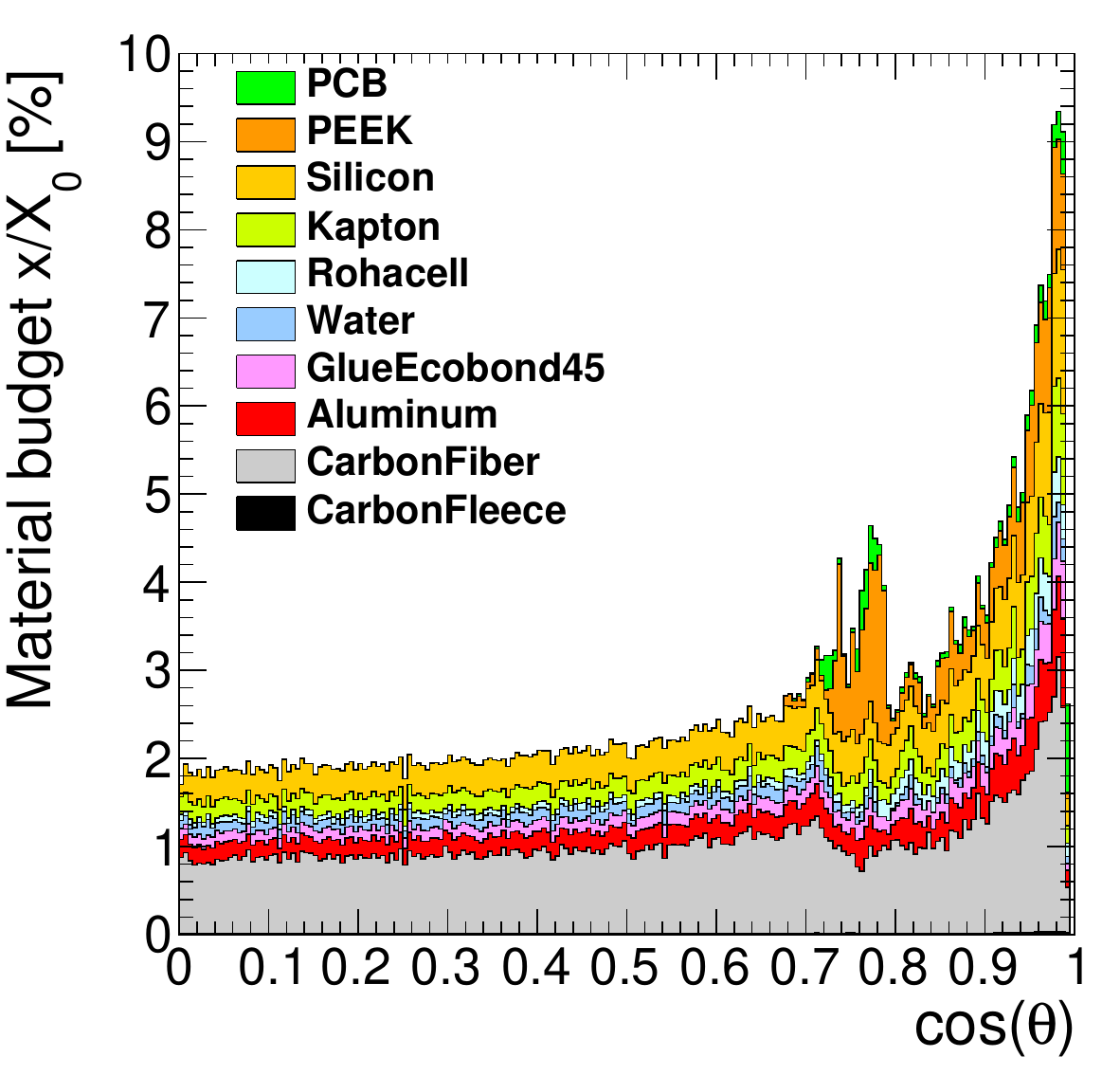}
    }
    \vspace*{-0.2cm}
    \caption{Material budget of the IDEA VXD DD4hep implementation.}
    \label{fig:VTX_material_budget}
    
\end{figure}

The resulting material budget of this implementation can be seen in 
\autoref{fig:VTX_material_budget}. In the first inner vertex barrel layer, the material budget at $\cos(\theta) = 0$ is around $\SI{0.25}{\percent}$ $X_0$ -- in line with the conceptual design report estimations. 
The most significant contributions to the material budget come from the CF support, followed by the silicon of the sensors. The material budget is not uniformly distributed in $\phi$ since the ladder support structures are not as wide as the ladder, and the ladders overlap to ensure hermeticity. This irregularity will lead to a dependence of the vertexing performance on $\phi$, which must be considered. The complete VXD features a material budget of around $\SI{2}{\percent}$ $X_0$, increasing when going to the forward region. The peak at $\cos(\theta) \sim 0.78$ comes from the outer vertex barrel structures that hold the ladders.

\begin{wrapfigure}[14]{r}{0.4\textwidth}
    \centering
    \vspace*{-0.0cm}
    \includegraphics[width=\linewidth,trim=0.5cm 0.4cm 0.2cm 0.4cm,clip]{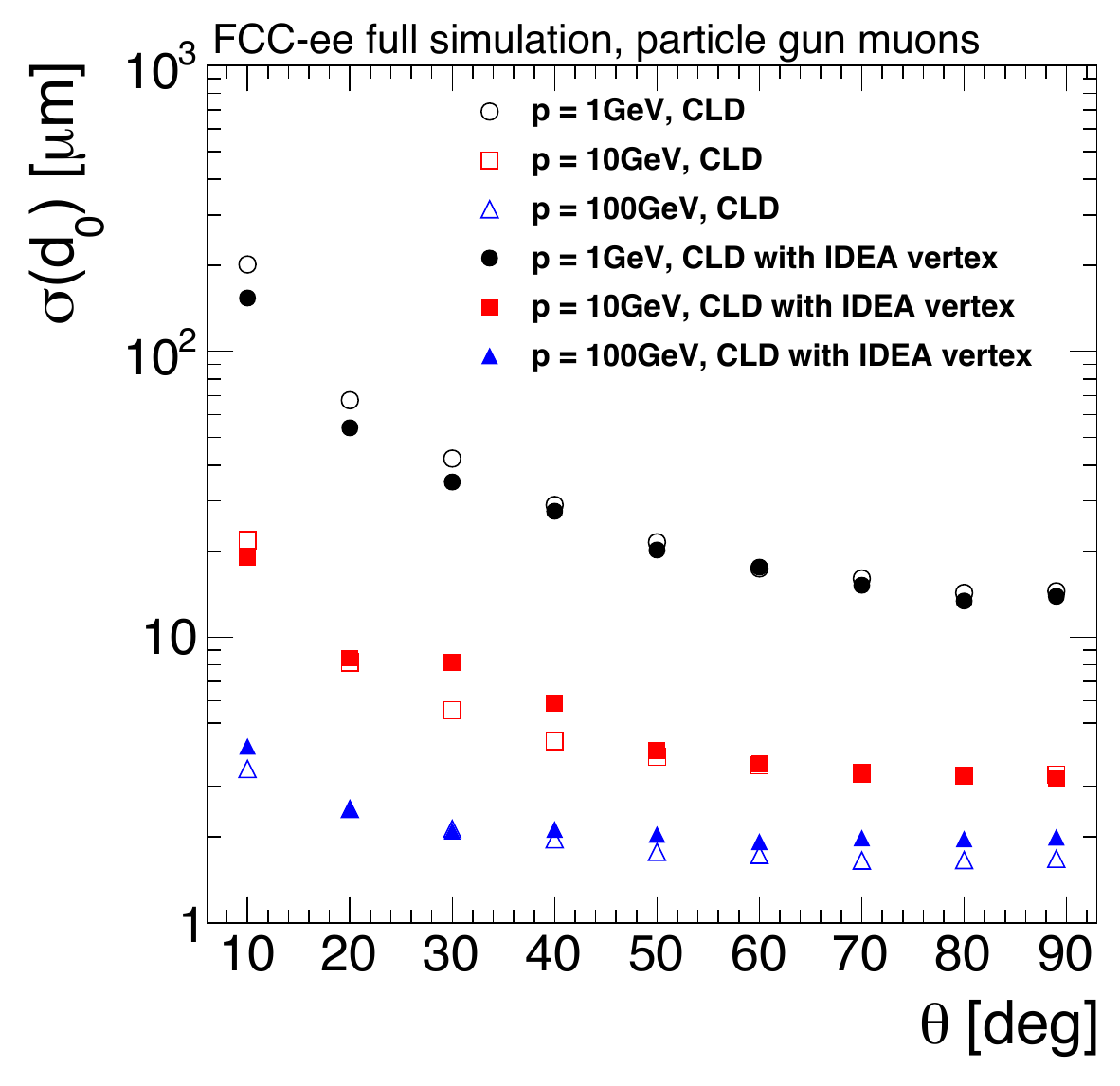}
    \vspace*{-0.8cm}
    \caption{Transverse impact parameter resolution of the IDEA VXD inserted into CLD compared to standard CLD.}
    \label{fig:vtx_performance}
\end{wrapfigure}

Since the IDEA drift chamber reconstruction was not available yet with the DD4hep model, the vertexing performance is assessed by inserting the IDEA VXD into CLD and removing one inner tracker barrel and two disks from there to avoid overlaps. Track and vertex reconstruction was performed using standard iLCSoft reconstruction \cite{Brondolin2020}. \autoref{fig:vtx_performance} shows the achieved $\sigma_{d_0}$ for different muon momenta and $\theta$ compared to CLD. The IDEA VXD has a better resolution for \SI{1}{GeV} muons thanks to the lower material budget of the single-layer design of IDEA compared to CLD as multiple Coulomb scattering has a greater effect on low momenta particles. 

CLD, however, is superior at higher momenta due to providing two hits per layer. The expected vertexing performance of IDEA is in line with the requirements of FCC-ee, but improving the vertexing capabilities could further enhance and extend the FCC-ee physics programme. A much-improved vertex resolution could, among others, enable the measurement of extremely rare $B^0$ decays to $K^{*0}$ and taus, for which only upper limits down to $\mathcal{B} (B^0 \rightarrow K^{*0} \tau^+ \tau^-) < \SI{3.1e-3}{}$ exist \cite{Dong2023}. The SM expectation is of $\mathcal{O}(10^{-7})$, but new physics can increase the branching ratio significantly.

\section{Ultra-light vertex detector for FCC-ee}

A previous Delphes fast simulation study showed that a VXD using thin, curved, wafer-scale MAPS similar to ALICE ITS3 would significantly improve the vertexing performance \cite{Freitag:2851362,Ilg2023}. This section describes the first full simulation geometry implementation of such a VXD concept.

ALICE ITS3 \cite{its3_tdr} uses the stitching technique to form wafer-scale sensors from multiple repeated sensor units (RSUs). Each ITS3 layer comprises two half-cylindrical sensors featuring ten RSUs in $z$ and three, four, or five in $\phi$ for the first, second and third layer. 

\begin{figure}[htbp]
    \vspace*{-0.2cm}
    \centering
    \subfloat[Complete first layer]{
        \includegraphics[width=0.43\textwidth]{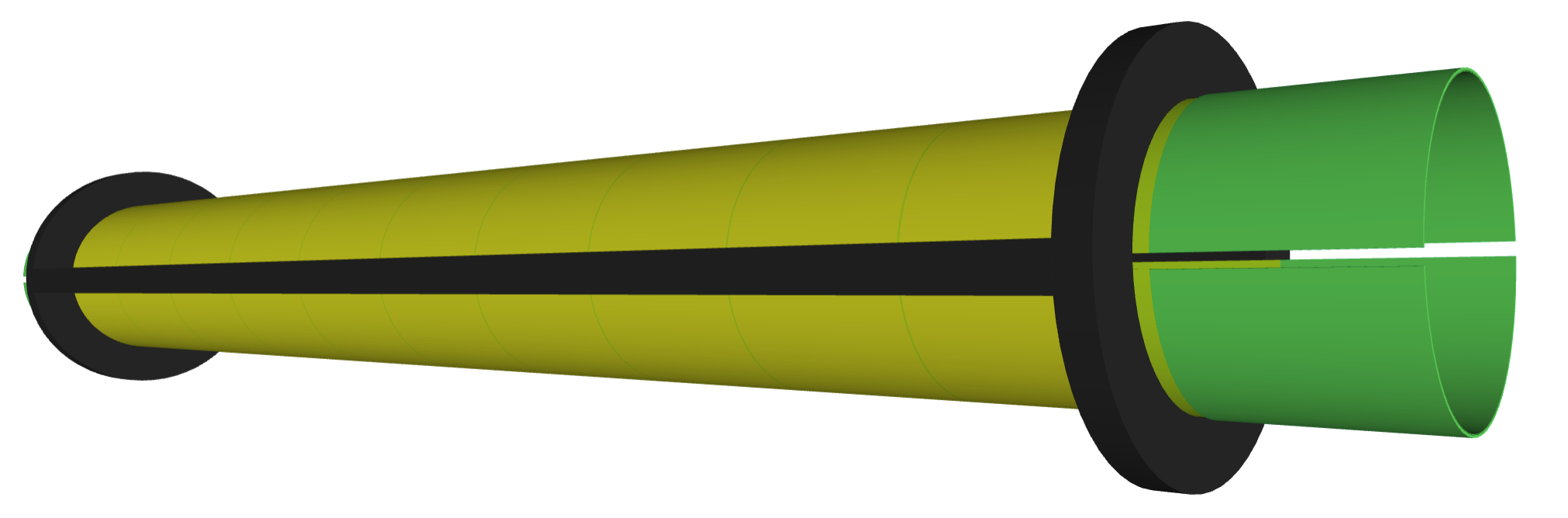}  
        \label{fig:vtx_ultra-light_L1}
    }
    \subfloat[Longitudinal cross-section of all four layers]{
        \includegraphics[width=0.57\textwidth]{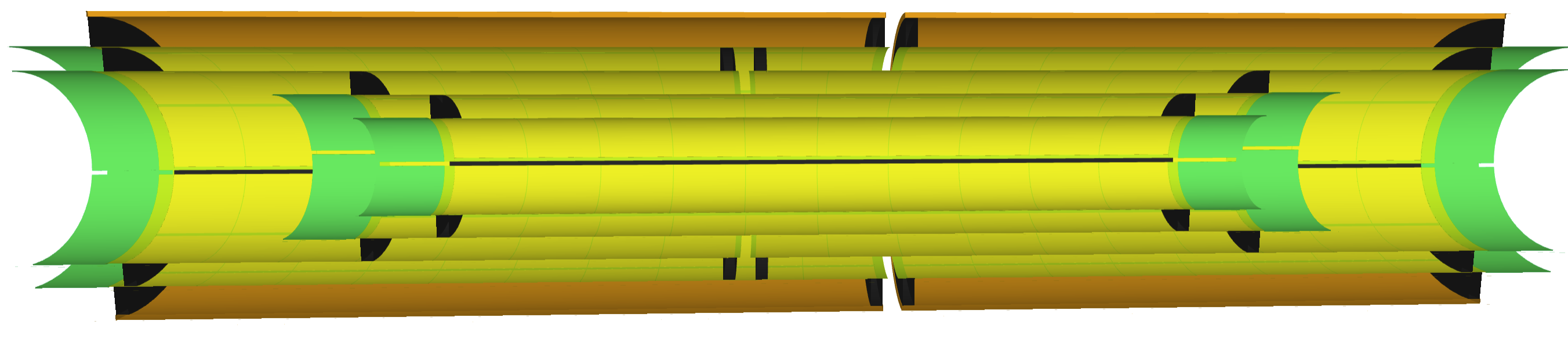} 
        \label{fig:vtx_ultra-light_tot}
    }
    \vspace*{-0.3cm}
    \caption{Curved wafer-scale sensors (yellow), carbon foam supports (black) and flex circuits (green) forming the ultra-light VXD concept implementation in DD4hep. }
    \label{fig:vtx_ultra-light}
\end{figure}

The ITS3 concept must be adapted to serve as an inner vertex detector for FCC-ee. The first layer, shown in Figure~\ref{fig:vtx_ultra-light_L1}, uses only two RSUs in $\phi$ to reach a radius of \SI{13.7}{mm} as in Section~\ref{sec:idea_vertex_design_mdi}, instead of \SI{18}{mm} in ITS3. The silicon sensors are assumed to have a thickness of $\SI{50}{\mu m}$, with an additional contribution of $\SI{16}{\mu m}$ silicon-equivalent for the metal layer connecting the RSUs. Only two carbon foam longerons and rings are foreseen to support the curved sensors. 

\begin{wrapfigure}[10]{r}{0.33\textwidth}
    \centering
    \vspace*{-0.6cm}
    \includegraphics[width=\linewidth]{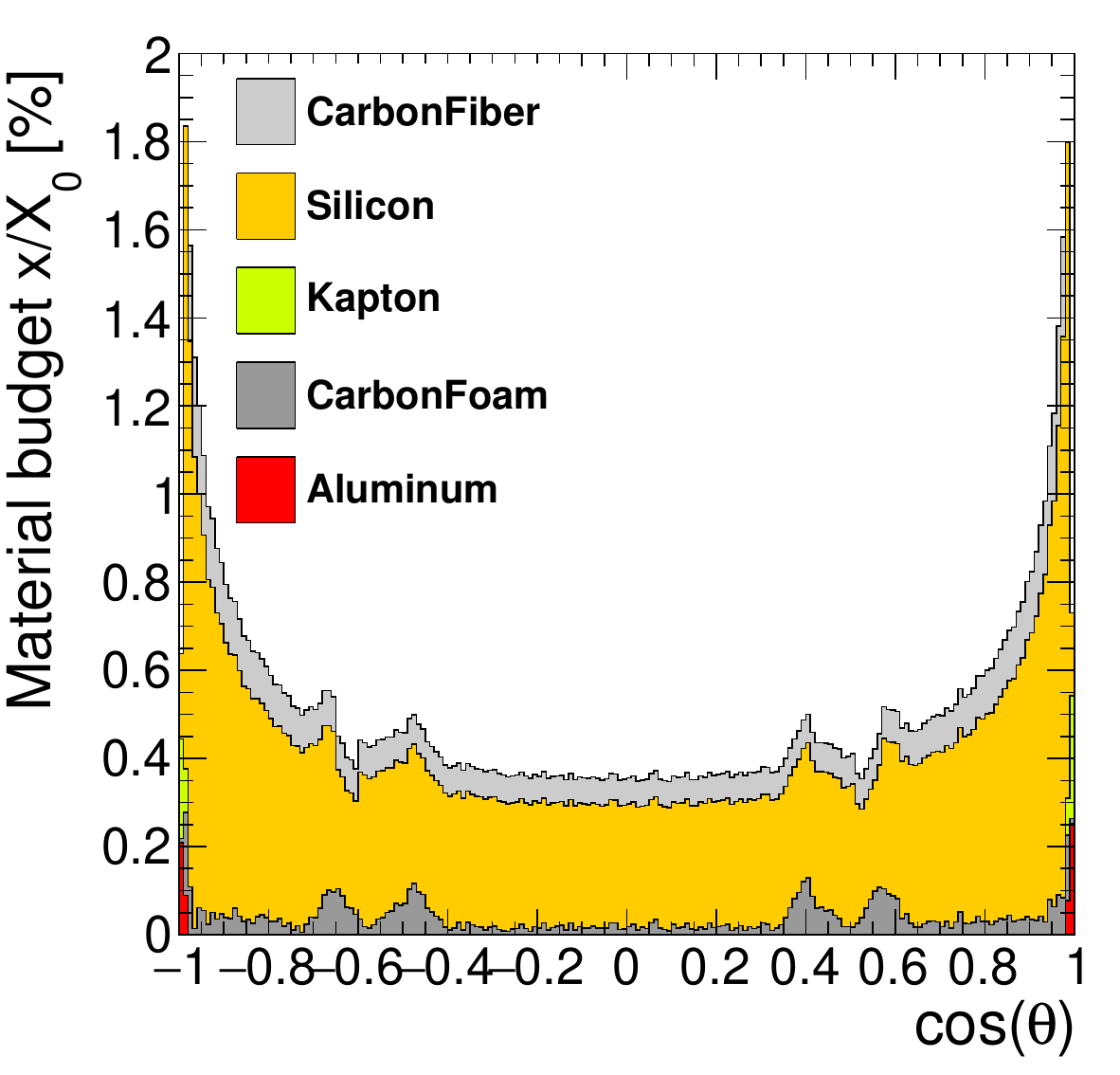}
    \label{fig:vtx_ultra-light_material_budget_VTXIB_curved_1D}
    \vspace*{-0.9cm}
    \caption{Material budget of the ultra-light inner VXD concept DD4hep implementation.}
    \label{fig:vtx_ultra-light_material_budget}
\end{wrapfigure}

The concept features four layers to compensate for the fact that not every layer detects every track due to gaps in $\phi$ acceptance of $\approx \SI{10}{\percent}$ from the RSU peripheries and gaps between the half-layers. Unlike in ITS3, the number of RSUs in $z$ increases for the outer layers for better forward coverage. For the third and fourth layers, the use of two sensors in $z$ per half-layer is foreseen to circumvent the limitation of the 12-inch wafer diameter. The third (fourth) layer uses 8 (10) RSUs on the $-z$ and 10 (8) on the $+z$ side, resulting in an asymmetric layout shown in Figure~\autoref{fig:vtx_ultra-light_tot}. Together with rotations of the layers in $\phi$, the gaps in acceptance of one layer are covered by the others.

\autoref{fig:vtx_ultra-light_material_budget} shows the material budget of the complete ultra-light inner VXD concept, amounting to only $\SI{0.35}{\percent}$ $X_0$. Excluding the external $\SI{200}{\mu m}$ thick CF tube for air-cooling, a single layer is $\approx \SI{0.075}{\percent}$ $X_0$ at $\cos(\theta) = 0$. This is about a factor three below the classic design ($\SI{0.25}{\percent}$, see Section~\ref{sec:idea_vertex_design_mdi}) and more uniformly distributed in $\phi$.

\section{Conclusions}

Precise vertex reconstruction will be crucial for many measurements at FCC-ee. The vertex detector of the IDEA detector concept is being engineered and simulated to establish the feasibility of the design.
The material budget is in line with previous estimations, and full simulation studies show that the achievable vertexing performance is similar to that of the CLD, featuring a better $d_0$ resolution for low-momenta particles while being worse for high momenta. 

Adapting curved, wafer-scale MAPS for FCC-ee vertex detectors promises to reduce the material budget of the innermost layers greatly. The first concept of such an FCC-ee vertex detector was designed and implemented in DD4hep. A reduction in the material budget of a factor $\approx 3$ is achieved, which could enable new measurements such as $B^0 \rightarrow K^{*0} \tau^+ \tau^-$. 

Moving to the pre-TDR phase of the FCC, aspects such as readout and system integration of the VXD will need to be studied extensively. \mycomment{\cite{Voutsinas2020,Boscolo_beamstrahlung}} Further R\&D on MAPS is also necessary since no single prototype fulfils all the stringent FCC-ee VXD requirements yet.



\bibliographystyle{JHEP}
\bibliography{biblio}
\end{document}